**Title:** A Neural Marker of Obsessive-Compulsive Disorder from Whole-Brain Functional Connectivity

**Authors and author addresses:**

Yu Takagi[1,2 †], Yuki Sakai[1,3 †], Giuseppe Lisi[1], Noriaki Yahata[4,5], Yoshinari Abe[3], Seiji Nishida[3], Takashi Nakamae[3], Jun Morimoto[1,2], Mitsuo Kawato[1,2], Jin Narumoto[3], Saori C Tanaka[1*]

[1]ATR Brain Information Communication Research Laboratory Group, Kyoto 619-0288, Japan

[2]Graduate School of Information Science, Nara Institute of Science and Technology, Nara 630-0192, Japan

[3]Department of Psychiatry, Graduate School of Medical Science, Kyoto Prefectural University of Medicine, Kyoto 602-8566, Japan

[4]Department of Youth Mental Health, Graduate School of Medicine, The University of Tokyo, Tokyo 113-0033, Japan

[5]Department of Molecular Imaging and Theranostics, National Institute of Radiological Sciences, National Institutes for Quantum and Radiological Science and Technology, Chiba 263-8555, Japan

[†]**Contributed equally to this work**

[*]**Corresponding author:** xsaori@atr.jp

**Abstract**

Obsessive-compulsive disorder (OCD) is a common psychiatric disorder with a lifetime prevalence of 2–3%. Recently, brain activity in the resting state is gathering attention for exploring altered functional connectivity in psychiatric disorders. Although previous resting-state functional magnetic resonance imaging studies investigated the neurobiological abnormalities of patients with OCD, there are concerns that should be addressed. One concern is the validity of the hypothesis employed. Most studies used seed-based analysis of the fronto-striatal circuit, despite the potential for abnormalities in other regions. A hypothesis-free study is a promising approach in such a case, while it requires researchers to handle a dataset with large dimensions. Another concern is the reliability of biomarkers derived from a single dataset, which may be influenced by cohort-specific features. Here, our machine learning algorithm identified an OCD biomarker that achieves high accuracy for an internal dataset (AUC = 0.81; N = 108) and demonstrates generalizability to an external dataset (AUC = 0.70; N = 28). Our biomarker was unaffected by medication status, and the functional networks contributing to the biomarker were distributed widely, including the frontoparietal and default mode networks. Our biomarker has the potential to deepen our understanding of OCD and to be applied clinically.

**Introduction**

Obsessive-compulsive disorder (OCD) is a common psychiatric disorder with a lifetime prevalence of 2–3%[1] that is characterized by obsessions (recurrent intrusive thoughts with excessive anxiety) and compulsions (excessive repetitive actions for reducing obsession-induced anxiety). Previous neuroimaging studies using structural and task-based functional magnetic resonance imaging (fMRI) have revealed neurobiological dysfunctions in OCD, most notably in the fronto-striatal circuit[2–6]. A meta-analysis of task-based fMRI studies using the symptom provocation paradigm revealed consistent increased activation within fronto-striatal regions[7]. Likewise, a multicenter voxel-based morphometric study revealed altered fronto-striatal gray and white matter volumes in patients with OCD[8]. Structural- or functional-MRI OCD classifiers constructed based on these findings have been reported[9–12]. Furthermore, methods for modulating the neural activity of the brain regions within the fronto-striatal circuit, such as deep brain stimulation (DBS), have been applied as clinical therapy for OCD[13].

Besides structural and task-based fMRI studies, resting-state fMRI (rs-fMRI) is gathering attention as a new means of exploring altered functional connectivity (FC) in OCD[14]. Several studies have reported that rs-fMRI can detect differences in FC between healthy controls (HCs) and patients with OCD[15–18], and find correlations with treatment response to medication[19] and behavioral therapy[20,21]. Furthermore, DBS reduced excessive FC within the fronto-striatal circuit, and the DBS-induced changes in FC and changes in symptom severity were correlated[22].

There are two types of rs-fMRI studies in OCD: hypothesis-driven, seed-based analyses and hypothesis-free, data-driven analyses. Many of the OCD rs-fMRI studies have used seed-based FC analyses with a focus on the hypothesis of local abnormalities, especially within the fronto-striatal circuit[15,16]. Recently, other data-driven studies have revealed more global abnormalities, involving a more complex combination of activity throughout the brain[23,24]. The latter approach does not require an *a priori* hypothesis; therefore, it has the potential to quantitatively evaluate the contribution of the fronto-striatal circuit relative to other brain regions[2].

Although previous rs-fMRI studies revealed neurobiological abnormalities in patients with OCD, the generalizability of these findings is still elusive. In fact, even for the most promising fronto-striatal circuit hypothesis, the findings have been inconsistent[15,16,24–26]; that is, the fronto-striatal circuit in OCD was hypoconnected in some studies[25,26] and hyperconnected in others[15,16,24]. Although these inconsistencies might be due to the clinical characteristics of each dataset[25,26], other studies suggested that the fronto-striatal circuit is not the only target should be considered. In other words, studies have suggested that there were abnormalities in addition to those observed in the fronto-striatal circuit, including the frontoparietal and default mode networks[5,27,28]. No study has quantitatively evaluated the relative importance of the fronto-striatal circuit relative to the whole brain. Thus, it is worth constructing a reliable FC-based "biomarker" that, with a subset of relevant FCs, allows the automatic distinction between patients with OCD and HCs. Such a biomarker may provide a novel framework in which psychiatric disorders, including OCD, are redefined biologically[29]. Only when such a biomarker is constructed through a fully data-driven approach, and when its validity is confirmed using independent cohorts of patients, will we be in a position to evaluate the relative importance of the fronto-striatal and other networks in the pathophysiology of OCD. Previously, an rs-fMRI study[23] attempted to predict the diagnosis of OCD in a data-driven and cross-validated manner, but the generalizability of the result was not verified using an external dataset. Indeed, it is quite challenging to construct a classifier with sufficient generalizability because of two major methodological difficulties[30]. First, the number of subjects in an rs-fMRI dataset is usually small relative to the high dimensionality of FCs. It is a well-known problem that applying a naïve machine learning classifier to such a dataset leads to over-fitting[31]. Second, findings obtained from a single dataset are heavily influenced by cohort-specific features such as sex, gender, and medication, that is, nuisance variables (NVs), which may lead to catastrophic over-fitting. The status of medication such as antidepressants and anxiolytics should also be included as NVs[23], so that the performance of the HC-OCD classifier is confirmed irrespective of medication status. Here, to overcome the aforementioned issues, we aimed to construct a reliable whole-brain rs-fMRI-based biomarker using a data-driven approach. The recently developed machine learning algorithm[31] employed a cascade of two statistical models: L1-norm regularized sparse canonical correlation analysis ($L_1$-SCCA) and sparse logistic regression (SLR)[32,33]. SLR can train a logistic regression model while objectively pruning features that are not useful for classifying OCD. For efficient training of SLR, however, the input dimension must be optimally reduced and the effects of NVs must be maximally removed. Therefore, before training SLR, feature selection was performed using $L_1$-SCCA, which can identify small sets of features that contribute to only a specific variable among the NVs. For example, we can identify a small set of features that are relevant for only the diagnosis label, but not relevant for the NVs such as age or medication use. By adopting a cascade of the sparse estimation methods, our procedure leads to sparse parameters with higher generalizability, while at the same time excluding features correlating with NVs. We hypothesized that our method could be used to distinguish patients with OCD from HCs, even in an external dataset. Furthermore, its predictions were unaffected by NVs such as medication status. Finally, we quantitatively evaluated the contribution of the fronto-striatal circuit relative to other brain regions for the classification of OCD.

## Results

**Constructing an rs-FC-based classifier**

All rs-fMRI data (N = 108) were collected at Kyoto Prefectural University of Medicine (KPUM), Kyoto, Japan. Fifty-six patients with OCD and 52 HCs were included. Table 1 summarizes the demographic data of the participants. There were 16 participants using medication (16 using antidepressants, 4 using antipsychotics, and 2 using anxiolytics). All patients were surveyed for obsessive symptoms using the Yale-Brown Obsessive-Compulsive Scale (Y-BOCS)[34].

Figure 1 shows the overview of the analysis. Pairwise, interregional FC was evaluated for each participant after standard preprocessing among 140 regions of interests (ROIs) covering the entire brain. The time courses of the voxels in each ROI were averaged to extract its time course. Then, for each participant, a matrix of FC between all ROIs was calculated by evaluating pairwise temporal Pearson correlations of the time course of blood oxygenation level-dependent signals. Further, to avoid multicollinearity between the input features, we used principal component analysis (PCA) and kept all obtained principal components (PCs).

This procedure enabled us to reduce the dimensionality of the input feature space from nearly 10,000 to the number of participants, thereby allowing the classifier to learn more stably. It should be noted that PCA was conducted using the whole training dataset; that is, an external dataset was not used to obtain the transformation matrix.

To avoid problems of over-fitting due to small sample size or irrelevant NVs, we applied the method developed in our previous study (see Methods)[31]. We constructed the classifier by combining two machine learning algorithms: SLR and $L_1$-SCCA[32,33]. In SLR, the probability distribution of the parameter vector is estimated by hierarchical Bayesian estimation, in which the prior distribution of each element of the parameter vector is represented as a Gaussian distribution. Based on the automatic relevance determination prior in the hierarchical Bayesian estimation method, irrelevant features are not used in the classification because the respective Gaussian prior distributions have a sharp peak at zero. In $L_1$-SCCA, latent relationships are identified between PCs and various attributes of each participant, including the diagnostic label and available demographic information. By selecting PCs connected to a canonical variable related to only the "Diagnosis" label and not to NVs, we aimed to reduce the interference of NVs. Here, we defined age, sex, handedness, and medication use (anxiolytics, antidepressants, or antipsychotics) as NVs. Our method avoids the problem of over-fitting by adopting a cascade of the sparse estimation method, a well-known approach for handling small sample sizes. Furthermore, it also avoids extracting irrelevant cohort-specific OCD features or NVs.

**Reliable classifier for OCD in the training set**

Leave-one-out cross-validation (LOOCV) was used to assess classification accuracy (see Methods). Participants with OCD could be separated from HCs with 73% accuracy and an area under the curve (AUC) of 0.81 (1,000-repetition permutation test, $P < 0.001$). Thus, the discriminatory ability of the classifier was high. The weighted linear summation (WLS or linear discriminant function) of the identified PC values of the classifier predicted the diagnostic label of each participant. Participants with a positive WLS were classified as OCD patients and those with a negative WLS as HCs. Figure 2a shows that the WLS distributions of the OCD and HC participants were separated to the right and left, respectively.

**Generalization of the classifier for the external dataset**

The generalizability of the classifier was tested by using an external dataset (N = 28) collected on a different MRI scanner from that used to collect the training dataset (see Methods). We used the same dataset as Sakai et al.[16] The patients were recruited at KPUM. None of the participants had been taking any kind of psychotropic medication for at least 8 weeks. Fifteen participants were entered into both experiments. In such a case, we used them in the training dataset and excluded them from the external dataset. Finally, 28 participants, including 10 patients with OCD, were used as the external dataset. Thus, there was no overlap between the training and external datasets. For this external dataset, the present classifier, trained with a different MRI scanner, performed well, with an AUC of 0.70 (1,000 repetitions permutation test, $P = 0.049$) (Fig. 2b). Notably, the external dataset was not involved in any part of classifier training. Therefore, the successful classification of the external dataset indicates that the developed biomarker has the ability to be generalized to a totally independent dataset.

**Effects of NVs and symptom severity**

Next, we investigated the effects of medication on classification accuracy. For the training set, the accuracy of LOOCV was 75% (12 of 16) and the AUC was 0.87 for patients on medication, and the accuracy of LOOCV was 67.5% (27 of 40) and the AUC was 0.79 for patients not on medication. Classification accuracy was not significantly different between the two populations (chi-squared test, $P = 0.581$). None of the patients in the external cohort were on medication. We also examined whether there were significant differences in age, sex, and symptom severity (Y-BOCS) between correctly and incorrectly classified patients. No significant difference was found in either the internal or external dataset in terms of age (two-sample t-test, $P > 0.05$), sex (chi-squared, $P > 0.05$), or Y-BOCS (two-sample t-test, $P > 0.05$).

**Contribution to the WLSs of each FC**

To understand how each FC contributed individually to the WLSs, their individual contributions to the WLSs through the selected PCs were calculated. As both PCA and the classifier are linear methods, the contribution of each FC can be calculated by examining the transformation matrix of PCA and the weight of the classifier. We considered 200 FCs that contributed the most to the WLSs. Figure 3a shows the spatial distribution of these 200 FCs that were identified from the dataset for the reliable classification of OCD and HC participants.

Next, to interpret their contributions in macroscale regions, all ROIs were grouped into 18 macroscale brain regions that were defined functionally in a previous study[35] (e.g., the default mode network[36]) and examined the number of FCs between each pair of regions in each network. The networks were defined according to the datasets in "BrainMap ICA," which identified intrinsic connectivity networks by applying independent component analysis (ICA) to BrainMap, a large-scale database of neuroimaging studies. Figure 3b shows the matrices for the 200 FCs in the macroscale regions. Diagonal and non-diagonal elements show within- and between-network FCs, respectively. Figure 3c shows a circle plot of the 200 FCs in the macroscale regions. The number of FCs in each of the two macroscale regions is presented as the thickness of the connection lines. Some trends were observed, for example, the right-lateral frontoparietal network contributed strongly relative to the other regions. However, the FCs were distributed widely rather than constrained locally. As for the FCs between the bilateral basal ganglia-thalamus and orbitofrontal cortex, only 2 FCs between the thalamus and orbitofrontal cortex were included among the 200 most contributing FCs (highlighted by the blue box in Fig. 3b). It is noteworthy that no FC between the orbitofrontal cortex and striatum was included in the 200 most contributing FCs. Note that, because this threshold (200 FCs) was arbitrary and not determined from the perspective of classification performance, we varied the threshold (100, 200, 400, and 600), and confirmed that the contribution of the FCs between the orbitofrontal cortex and striatum was still small. That is, only 0, 0, 1, and 1 FCs between the orbitofrontal cortex and striatum were selected for each respective threshold.

**Discussion**

A reliable neuroimaging-based classifier for OCD was developed in this study by investigating whole-brain FC patterns using rs-fMRI data. This classifier incorporated the PCs of FCs distributed across the brain, and achieved a high AUC of 0.81 with an accuracy of 73%. Further, the classifier could be generalized to an external dataset (AUC of 0.70). To our knowledge, no neuroimaging-based classifier for OCD has been shown to be generalizable using an external, independent dataset. By interpreting the classifier, we evaluated for the first time the relative contribution of the fronto-striatal and other networks in the successful classification of OCD.

We found that the FCs contributing to the classification were distributed widely rather than being locally constrained. Specifically, many of the FCs were involved in the frontoparietal or default mode network. It is noteworthy that there were relatively fewer investigations that focused on the frontoparietal and default mode networks rather than the fronto-striatal circuit. However, both seed-based[17,27,28] and data-driven studies[21,23,37] have reported abnormalities of these networks besides the fronto-striatal circuit. Intriguingly, frontoparietal abnormalities driving OCD pathophysiology were also suggested by a previous study[38]. Although this previous study found DBS-induced changes in the fronto-striatal circuit, our result suggests that DBS could trigger broad changes in the FC patterns of the whole brain. Our findings suggest that, although previous studies, including our laboratory's, have often reported abnormalities in the fronto-striatal circuits[16], other networks should also be examined in the investigation of OCD.

Although the successful construction of a structural- or functional-MRI OCD classifier has been reported previously[9–12], the present study is the first to classify OCD across internal and external datasets. This was achieved because our analysis pipeline was fully data-driven and cross-validated, instead of using the seed-based analysis employed in most of the previous studies. Furthermore, we employed a cascade of sparse estimation methods by using $L_1$-SCCA and SLR[31]. We were able to avoid the over-fitting problem by extracting optimal PCs that were relevant only to the core OCD characteristics. At the same time, we could eliminate the effects of NVs such as age, sex, and medication in the feature selection process. Specifically, we did not observe a clear difference in classification accuracy between patients with and without medication. Medication reportedly significantly affects rs-FC patterns[39], and a naïve algorithm might over-fit the difference induced by the medication use, which leads to a reduction of generalization accuracy for non-medicated OCD patients in validated data.

The output of the OCD classifier might provide a reliable measure of an individual's "OCD-ness" along one of the biological dimensions in psychiatric disorders, because our OCD classifier was successfully generalized to an external dataset for the first time. In the field of psychiatry, we have been unable to find any neuroscientific evidence to support the breakdown of complex psychiatric disorders into separate categories. Therefore, the hypothesis of a multiple psychiatric disorder spectrum is gaining attention[40]. According to this view, psychiatric disorders are the product of shared risk factors, or dimensions, that lead to abnormalities. Although the findings from brain imaging[31,41] and genetic studies[42] support this idea, this hypothesis is still premature because of the scarcity of reliable dimensions. Our OCD classifier provides a biologically defined continuous index of OCD, and it can reliably separate HCs from patients with OCD. Therefore, it can be considered an objective, reliable dimension for the spectrum. Although the outputs of the biomarker were not related to a conventional clinical severity-index, our machine learning-based biomarker can be an objective and reliable complementary measure, given that the current diagnosis for this psychiatric disorder is based on a subjective report from an individual. Further studies evaluating the relationship between the classifiers of multiple psychiatric disorders are needed for a deeper understanding of psychiatric disorders and for clinical application.

A limitation of the present study is that we cannot directly compare our finding with previous studies investigating local brain regions[15,16]. This is because we employed the PCs of FCs, and they represent a linear combination of whole brain FCs. This is the conventional approach in the field of machine learning to avoid the over-fitting problem when using a dataset with a small sample size and high dimensionality. In addition, unlike our previous study that employed a multi-site dataset[31], all participants in the training dataset were scanned in the same site. It might also lead to difficulties with generalization at another site without PCA because of the presence of uncontrolled site-specific NVs. A future study with a much larger sample size assessed at multiple sites will investigate the contribution of each FC independently. Furthermore, in removing artifacts, we applied motion correction, NV regression, and scrubbing. Although recent studies have proposed alternative procedures for denoising (e.g., Kundu et al.[43]), there is still debate about the optimal denoising procedure. The development of a methodologically more appropriate alternative may lead to further improvement of classification accuracy. In addition, although classification accuracy was not significantly different (chi-squared test, $P = 0.100$) between the internal and external datasets, our biomarker worked better for LOOCV in the internal dataset than for the generalization to the external dataset. This is reasonable because, usually, generalization to a different dataset is more difficult than to the same dataset due to the distribution of samples or measurement noise being different from those of the training dataset. Finally, it should be noted that the sample size of the external dataset was not large (N = 28). Even though our biomarker was also verified in a fully cross-validated manner with the much larger internal dataset (N = 108), its generalizability might not be estimated with high accuracy.

In summary, we have developed the first generalizable rs-fMRI-based classifier for OCD. It reliably distinguished participants with OCD from HCs even in an independent validation dataset. Our whole-brain biomarker may shed light on the neural substrates of OCD in the form of the abnormal FC pattern across the whole brain.

## Methods

**Training dataset used for the construction of the OCD classifier**

*Participants:* All resting state fMRI data (N = 108) were collected at KPUM; 69 of these participants were also included in the study of Abe et al.[44] Although a few studies have tried to construct a biomarker in a fully cross-validated manner using machine learning, our sample size is larger than that of a previous study of an OCD biomarker[23] (N = 46). The demographic data for all experiments are shown in Table 1. Patients with OCD were recruited at KPUM. Trained, experienced clinical psychiatrists and psychologists assessed all participants. All patients were primarily diagnosed using the Structured Clinical Interview for DSM-IV Axis I Disorders-Patient Edition (SCID)[45]. Exclusion criteria were 1) cardiac pacemaker or other metallic implants or artifacts; 2) significant disease, including neurological diseases, disorders of the pulmonary, cardiac, renal, hepatic, or endocrine systems, or metabolic disorders; 3) prior psychosurgery; 4) DSM-IV diagnosis of mental retardation and pervasive developmental disorders based on a clinical interview and psychosocial history; and 5) pregnancy. We excluded patients with a current DSM-IV Axis I diagnosis of any significant psychiatric illness except OCD as much as possible, and only 1 patient with trichotillomania, 1 patient with tic disorder and specific phobia, and 1 patient with bulimia nervosa were included as patients with a comorbid condition. There was no history of psychiatric illness in the control group as determined by the SCID-Non-Patient Edition[46]. In addition, they reported no history of psychiatric treatment in any of their first-degree relatives. Handedness was classified based on a modified 25-item version of the Edinburgh Inventory. The Medical Committee on Human Studies at KPUM approved all procedures in this study. All participants gave written, informed consent after receiving a complete description of the study. All methods were carried out in accordance with the approved guidelines and regulations.

*Image acquisition:* A whole-body 3-T MR system (Achieva 3.0T Quasar Gyroscan Intera; Philips Medical Systems, Best, The Netherlands) with an 8-channel phased-array head coil at the Kajiicho Medical Imaging Center was used to generate magnetic resonance images. Functional data were collected using gradient echo planar imaging (EPI) sequences (echo time/repetition time, 30/2000 ms; flip angle, 80°; field of view, 192 mm$^2$; imaging matrix, 64

× 64, 39 slices; slice thickness, 3.0 mm, no gaps). High-resolution (1.0 × 1.0 × 1.0 mm) T1-weighted magnetization-prepared rapid gradient echo images were also acquired before scanning the functional data. The first 6 (additional) images were discarded to allow magnetization to reach equilibrium. All participants underwent an approximately 6 min and 40 s resting-state scan, resulting in a total of 200 volumes. They were instructed simply to keep their eyes closed, not to think of anything, and not to fall asleep.

**External validation dataset**

*Participants:* We used the same dataset as Sakai et al.[16] Fifteen participants were also included in the training dataset; therefore, they were excluded from the validation dataset. Finally, 28 participants were used as the external validation dataset. Thus, there was no overlap between the training and external validation datasets. Patients with a current DSM-IV Axis I diagnosis of any significant psychiatric illness except OCD were excluded. The other settings were the same as for the training dataset. The Medical Committee on Human Studies at KPUM approved all procedures in the study. All participants gave written, informed consent after receiving a complete description of the study. All methods were carried out in accordance with the approved guidelines and regulations.

*Image acquisition:* A whole-body 1.5-T MR system (Gyroscan Intera; Philips Medical Systems, Best, The Netherlands) with a 6-channel phased-array head coil was used to generate magnetic resonance images. Foam pads were used to reduce head motion and scanner noise. Functional data were collected using gradient EPI sequences (echo time/repetition time, 40/2411 ms; flip angle, 80°; field of view, 192 mm$^2$; imaging matrix, 64 × 64, 35 slices; slice thickness, 3.6 mm, no gaps). High-resolution (1 × 1 × 1.5 mm) T1-weighted magnetization-prepared rapid gradient echo images were acquired before each resting image. All participants underwent an approximately 8 min resting-state scan, resulting in a total of 200 volumes. The experimental settings for the resting-state scan were the same as for the training dataset.

**Preprocessing**

We used a preprocessing method similar to that of Yahata et al.[31] for both the training and external datasets. We used Statistical Parametric Mapping 8 (Wellcome Trust Centre for Neuroimaging, London, UK; http://www.fil.ion.ucl.ac.uk/spm/software/) in MATLAB (The MathWorks, Inc., Natick, MA) for preprocessing and statistical analyses. First, head motion was compensated for by collecting raw functional images for slice-timing and realigning them to the mean image of that sequence. Second, the structural images were co-registered to the mean functional image and segmented into 3 tissue classes in Montreal Neurological Institute (MNI) space. Using associated parameters, we normalized the functional images and resampled them in a 2 × 2 × 2 mm grid. Third, the images were smoothed by a Gaussian function with a full width at half-maximum of 6 mm. To avoid the effects of motion artifacts, the pre-processed sequence of functional images was examined as follows. First, the mean relative displacement in each of the 6 motion parameters (translation along and rotation with respect to the x, y, and z axes) was evaluated by calculating the mean of the absolute frame-to-frame relative changes in each parameter through a given time series (namely, the mean of $|\Delta_p(i)| = |p_{i+1} - p_i|$ across the time series, where $p$ is one of the 6 motion parameters and $i$ specifies the time point). In both the training and external datasets, no statistically significant difference between the groups was noted in this measure for the 6 motion parameters (two-sample t-test, $P > 0.05$ for all parameters in both datasets). Next, frame displacement (FD) was calculated for each participant at each time point by summing all 6 parameters. Using this FD, we used the "scrubbing" procedure to identify and exclude any frame affected by excessive head motion[47]. Specifically, a frame was flagged and removed, along with the previous and two subsequent frames, from correlation analysis, if the associated FD exceeded 0.5 mm. For both datasets, there was no difference in the number of frames that passed this procedure between the HC and OCD populations (two-sample t-test, $P > 0.05$).

**Interregional correlation analysis**

Pairwise, interregional FC was evaluated for each participant among 140 ROIs covering the entire brain. Each region's spatial extent was defined anatomically according to the digital atlas of the BrainVISA Sulci Atlas (BSA)[48]. As this atlas does not include the cerebellum, the 3 subregions of the cerebellum were appended to it based on the anatomical automatic labeling (AAL) package[49]. This BSA-AAL composite atlas was resampled in $2 \times 2 \times 2$ mm grid MNI space. The time course of the voxels in each region was averaged to extract its representative time course. Further, we excluded ROIs with zero-variance in at least 1 participant. The time course sets were band-pass filtered (0.008–0.1 Hz) prior to the following regression procedure. The filtered time courses were linearly regressed by the temporal fluctuations of the white matter, cerebrospinal fluid, and entire brain as well as the 6 head motion parameters. The fluctuation in each tissue class was determined from the average time course of the voxels within a mask created by the segmentation procedure of the T1 image. The mask for the white matter was eroded by 1 voxel to consider a partial volume effect. These extracted time courses were band-pass filtered (0.008–0.1 Hz) before linear regression, as was performed for the regional time courses. Then, for each participant, a matrix of FCs between all ROIs was calculated while discarding flagged frames, if any, in the previous procedure (scrubbing). The scrubbing procedure removed any frames exhibiting abrupt head movements that could be the source of high-frequency fluctuations in the filtered time course[50]. The FC matrices are symmetric, so values on only one side of the diagonal were kept, resulting in the number of samples × number of FC matrices. Further, to reduce the dimensionality of the matrix from nearly 10,000 to the number of participants, we used PCA and kept all obtained PCs for the following classification analyses, resulting in the number of samples × number of PC matrices. This procedure allowed the classifier to avoid multicollinearity between the input features and to learn in a stable manner. PCA was conducted using the whole training dataset.

**Selecting FCs as the OCD classifier**

To avoid the problems of over-fitting because of small sample size or irrelevant NVs, we applied the method developed by Yahata et al.[31] The procedure for selecting relevant PCs, training the predictive model, and assessing its generalization ability was performed as a sequential process of nested feature-selection and LOOCV. In each LOOCV fold, all-but-one participant was used to train the SLR classifier, while the remaining participants were used for evaluation. SLR can train a logistic regression model while objectively pruning PCs that are not useful for classifying OCD. For efficient training of SLR, however, the input dimension must be optimally reduced and the effects of NVs must be maximally removed. Therefore, before LOOCV, nested feature selection was performed using $L_1$-SCCA.

**Prediction of the diagnostic label**

Logistic regression analysis was used as the classifier to diagnostically label from the identified PCs. A logistic function was used to define the probability of a participant belonging to the OCD class:

$$P(y=1|\hat{\mathbf{z}};\mathbf{w}) = \frac{1}{1+\exp(-\mathbf{w}^T\hat{\mathbf{z}})}$$

here, $y$ is the diagnosis class label (OCD, $y = 1$; HC, $y = 0$); $\hat{\mathbf{z}} = [\mathbf{z}^T, 1]^T \in \mathbb{R}^{m+1}$ is a feature vector with an augmented input, where the feature vector $\mathbf{z}$ is the PCs of a participant's rs-fMRI sample. Using the augmented input "1" is a standard approach to introduce constant (bias) input for the classifier. $\mathbf{w} \in \mathbb{R}^{m+1}$ is the weight vector of the logistic function. To decrease the dimension of the feature vector further, which was already reduced by $L_1$-SCCA according to the equation, we used an SLR method. SLR automatically selects OCD-classification-related features as input for the logistic function.

**Author contributions:**

Y.T., Y.S., M.K, J.N., and S.C.T. conceived and designed the study; Y.S., Y.A., S.N., and T.N. recruited the participants of the study and collected their clinical and imaging data; Y.T., Y.S., L.G., N.Y., and J.M. analyzed the data; Y.T., Y.S., G.L., N.Y., Y.A., S.N., T.N., J.M., M.K., J.N., and S.C.T. wrote the manuscript. All authors reviewed the manuscript.

**Acknowledgements:**

This study is the result of "Brain Machine Interface Development," carried out under the Japan Agency for Medical Research and Development. We thank T. Okada, H. Ito, and technical engineers for their assistance in MRI data acquisition. We thank K. Tamura, S. Kimura, and K. Inoue for their assistance in the assessment of the patients. We thank the Joint Usage/Research Center at ISER, Osaka University for financial support.

**Competing financial interests**:

The authors declare no competing financial interests.

**Figure/Table Legends**

**Figure 1: Schematic diagram of the procedure for selecting FCs as an OCD biomarker and assessing their predictive power.**

A sketch map of prediction analysis. Resting state functional connectivity (rs-FC) matrices were processed through the cascading feature selection procedure. Left-out participants and all participants in the external validation dataset were classified based on the classifier derived from the rs-FC matrix from the other participants.

**Figure 2: Distribution of weighted linear summations (WLSs) of functional connections used for the classification of the OCD and HC populations.**

(**a**) The number of HC (white) and OCD (black) participants in the internal dataset in a specific WLS interval of width 5 is shown as a histogram. (**b**) WLS for the validation dataset in a specific WLS interval of width 2 is shown as a histogram.

**Figure 3: Functional connections used in the classification of the OCD and HC populations.**

(**a**) The 200 most contributing FCs from the left (left top), posterior (left bottom), and top (right) to the WLSs are visualized. (**b**) Matrices for the most contributing 200 FCs in 18 macroscale regions that were functionally defined in a previous study[35]. Diagonal and non-diagonal elements show within- and between-network FCs, respectively. The blue box highlights the corresponding area in the matrix discussed in the main text, i.e., FC between the orbitofrontal and basal ganglia-thalamus networks. The color bar indicates the number of FCs included between two networks. (**c**) Circle plot of the 200 most contributing 200 FCs in 18 macroscale regions. The number of FCs in each of the 2 macroscale regions is presented as the thickness of the connection lines (edges). The connections within the same network are colored blue, and connections between two different networks are colored red.

**Table 1: Demographic information of the participants used to construct the classifier for the OCD and HC populations (mean ± standard deviation).** All demographic

distributions are matched between the OCD and HC populations in the internal and external datasets ($P > 0.05$).

NR, not recorded.

NA, not applicable.

**Table 1**

|  | Internal | | External | |
|---|---|---|---|---|
|  | OCD | HC | OCD | HC |
| Male/Female | 23/33 | 26/26 | 4/6 | 8/10 |
| Age (years) | 32.64 ± 9.63 | 29.40 ± 7.46 | 31.60 ± 10.36 | 29.89 ± 8.69 |
| Handedness (R/L) | 51/5 | 50/2 | 9/1 | 15/3 |
| Y-BOCS | 21.13 ± 6.32 | NR | 23.8 ± 5.77 | NR |
| Medication (yes/no) | 16/40 | NA | 0/10 | NA |

# Figure 1

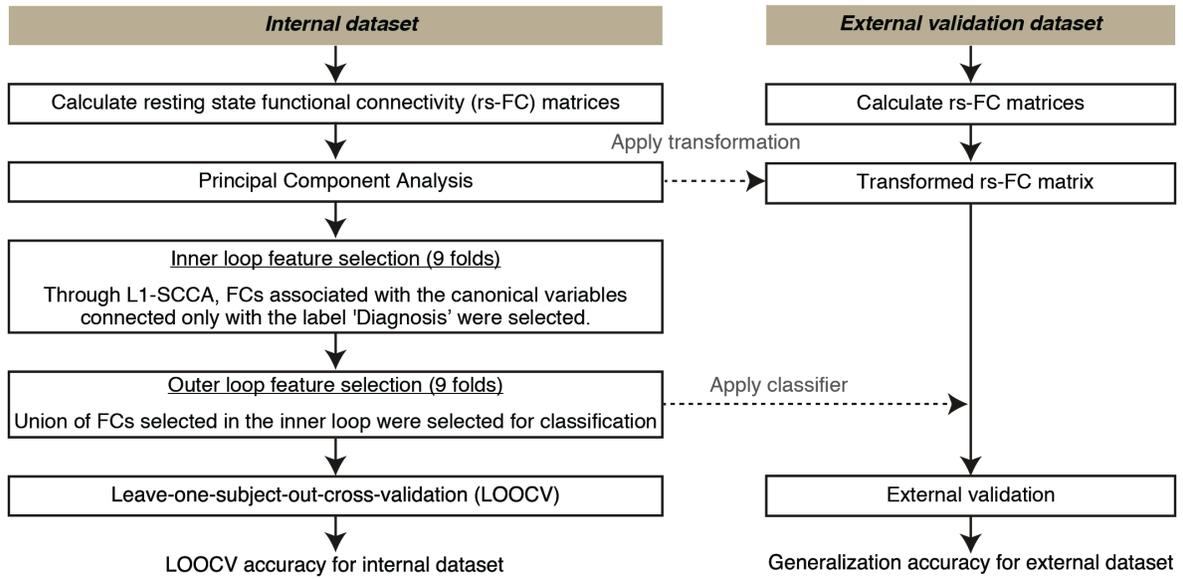

**Figure 2**

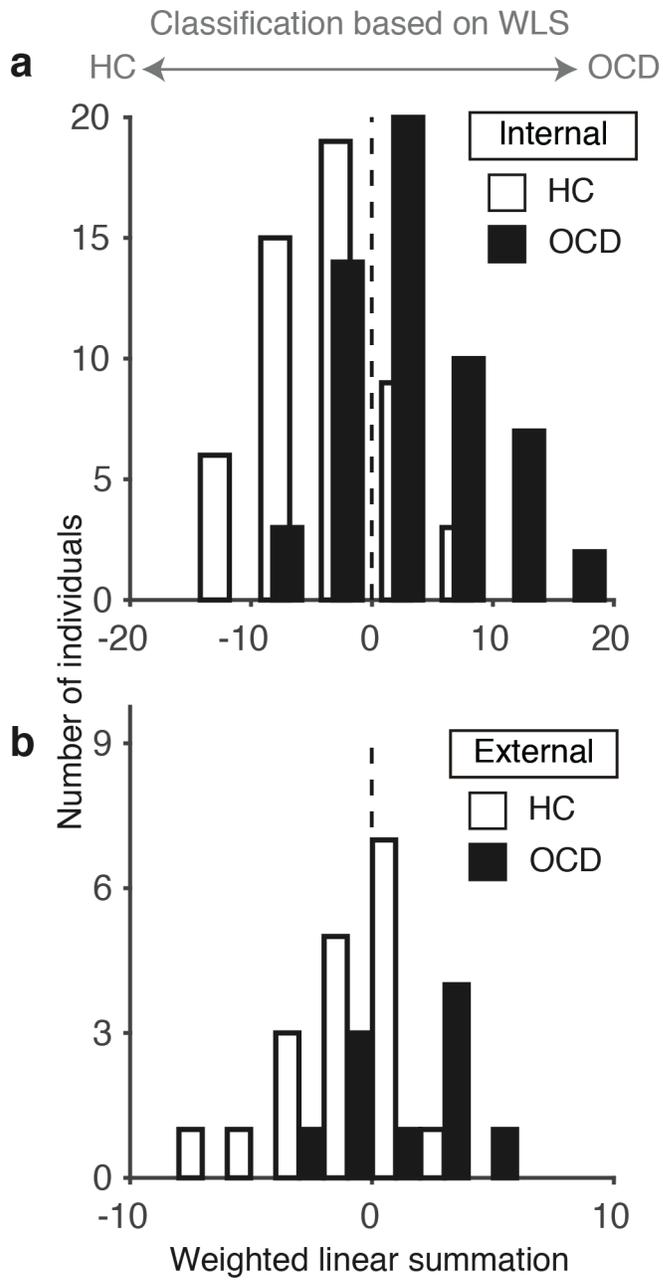

# Figure 3

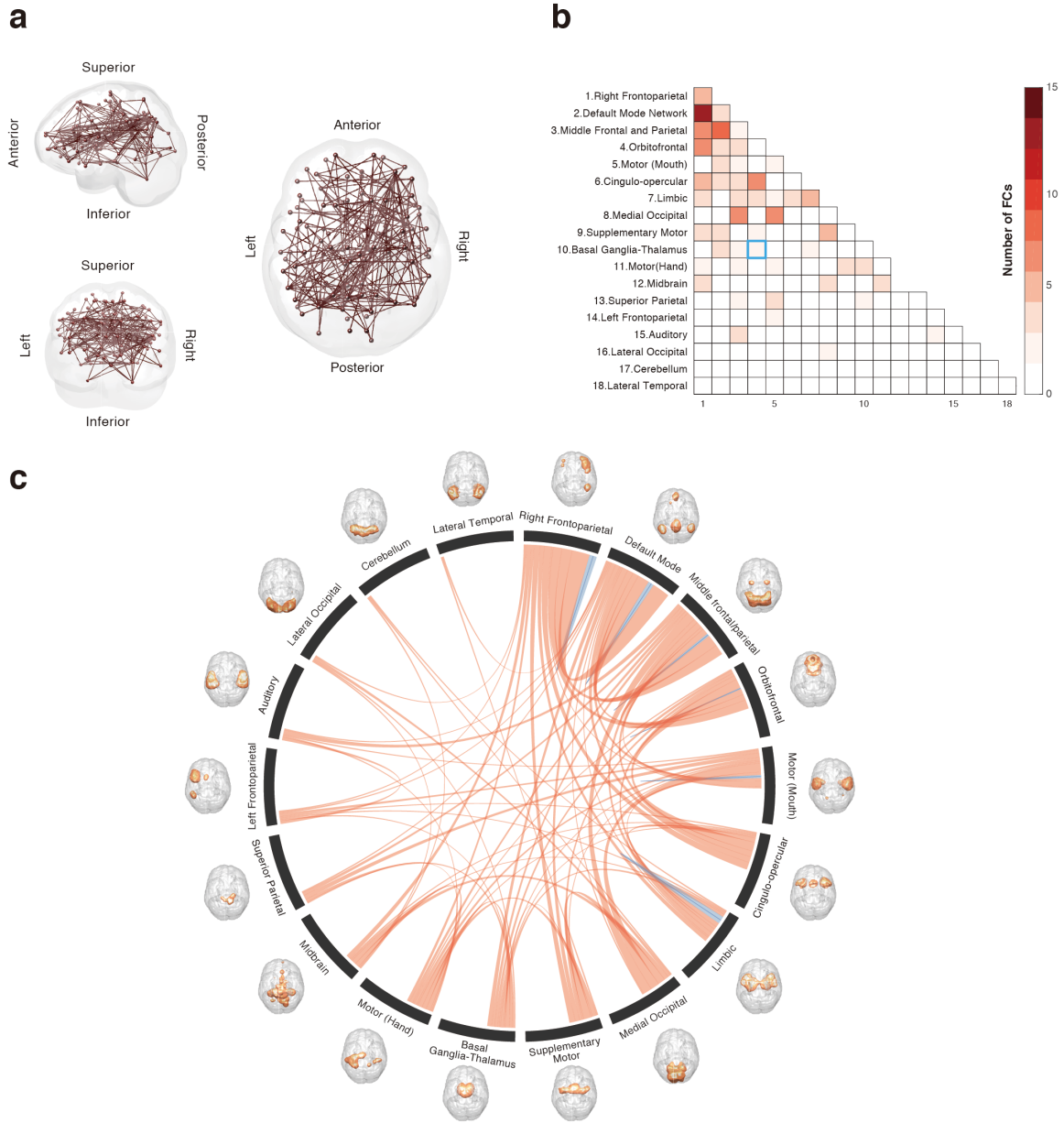